\documentclass[12pt]{article}

\usepackage{graphicx}
\usepackage{amsmath, amssymb}

\begin{document}
  
\begin{titlepage}

\begin{center}

\hfill TU-782 \\
\hfill October, 2007

\vspace{0.5cm}
{\large\bf  Gravitino Dark Matter\\
with Weak-Scale Right-Handed Sneutrino}

\vspace{1cm}
{\bf Koji Ishiwata}$^{(a),}$
\footnote{E-mail: ishiwata@tuhep.phys.tohoku.ac.jp},
{\bf Shigeki Matsumoto}$^{(a, b),}$
\footnote{E-mail: smatsu@tuhep.phys.tohoku.ac.jp},
and
{\bf Takeo Moroi}$^{(a),}$
\footnote{E-mail: moroi@tuhep.phys.tohoku.ac.jp}

\vspace{1cm}
{\it $^{(a)}${Department of Physics, Tohoku University,
              Sendai 980-8578, Japan}
\\
~
\\
     $^{(b)}${Tohoku University
              International Advanced Research and Education Organization,
              Institute for International Advanced Interdisciplinary Research,
              \\
              Sendai, Miyagi 980-8578, Japan} }

\vspace{1cm}
\abstract{ We consider cosmological implications of supersymmetric
  models with right-handed (s)neutrinos where the neutrino masses are
  purely Dirac-type.  We pay particular attention to the case where
  gravitino is the lightest superparticle while one of the
  right-handed sneutrinos is next-to-the-lightest superparticle.  We
  study constraints from big-bang nuleosynthesis and show that the
  constraints could be relaxed compared to the case without
  right-handed sneutrinos.  As a result, the gravitino-dark-matter
  scenario becomes viable with relatively large value of the
  gravitino mass.  We also discuss constraints from the structure
  formation; in our model, the free-streaming length of the gravitino
  dark matter may be as long as $O(1\ {\rm Mpc})$, which is comparable
  to the present observational upper bound on the scale of
  free-streaming.  }

\end{center}
\end{titlepage}
\setcounter{footnote}{0}

\section{Introduction}
\label{sec:intro}

Existence of dark matter in our universe, which is strongly supported
by a lot of recent cosmological observations
\cite{WMAP1,Seljak:2004xh,Spergel:2006hy}, requires physics beyond the
standard model. This is because there is no viable candidate for dark
matter in the particle content of the standard model. Many
possibilities of dark matter have been discussed in various frameworks
of particle physics models so far \cite{ReviewDM}. Importantly,
properties of the dark matter particle depend strongly on the particle
physics model we consider.

In this article, we adopt supersymmetry (SUSY) as new physics
beyond the standard model.  In such a case, probably the most popular
candidate for dark matter is thermally produced lightest neutralino
which is usually assumed to be the lightest superparticle (LSP); in
some part of the parameter space of the minimal supersymmetric
standard model (MSSM), the relic density of the lightest neutralino
well agrees with the present mass density of dark matter
observed. However, as we discuss in the following, the lightest
neutralino is not the only possibility of dark matter in
supersymmetric models.

If we try to build a supersymmetric model which accommodates with all
theoretical and experimental requirements, we expect that there exist
new exotic particles which are not superpartners of the standard model
particles.  For example, if local SUSY is realized in nature,
gravitino $\psi_\mu$ which is the superpartner of the graviton should
exist. In addition, superpartners of right-handed neutrinos, which are
strongly motivated to explain neutrino masses indicated by the
neutrino oscillation experiments, may also exist. In particular, if
the neutrino masses are Dirac-type, superpartners of the right-handed
neutrinos are expected to be as light as other MSSM superparticles in
the framework of gravity-mediated SUSY breaking. Importantly, one of
those new particles may be the lightest superparticle and hence may be
dark matter. In addition, existence of these exotic superparticles may
significantly change the phenomenology of dark matter in
supersymmetric models.

In this paper, we consider the supersymmetric model in which the
neutrino masses are Dirac-type and discuss cosmological implications
of such a scenario.  The possibility of the right-handed sneutrino LSP
has already been discussed in Ref.\cite{AsakaIshiwataMoroi}; it has
been pointed out that, if one of the right-handed sneutrinos is the
LSP, the present relic density of the right-handed sneutrino
$\tilde{\nu}_R$ may be as large as the dark matter density and hence
the scenario of $\tilde{\nu}_R$ dark matter can be realized. Here, we
consider another case where the gravitino is the LSP and one of the
right-handed sneutrinos is the next-to-the-lightest superparticle
(NLSP). If the gravitino is the LSP, it may be a viable candidate for
dark matter also in the case without the right-handed sneutrinos
\cite{Moroi:1993mb,FengRajaramanTakayama,Ellis:2003dn}. In such a
case, however, stringent constraints on the scenario are obtained from
the study of the gravitino production at the time of the reheating
after inflation and also from the study of the big-bang
nucleosynthesis (BBN) reactions. With the right-handed sneutrino NLSP,
we reconsider cosmological constraints on the gravitino LSP
scenario. We pay particular attention to the BBN constraints and also
to the constraints from the structure formation of the universe. We
will see that the BBN constraints could be significantly relaxed if
there exists the right-handed sneutrino NLSP. In addition, we will
also see that the free-streaming length of the gravitino dark matter
may be a few Mpc, which is as long as the present sensitivity to the
free-streaming length from observations.  Thus, detailed understanding
about the mechanism of structure formation has important impact on our
scenario.

The organization of this paper is as follows.  In the next section, we
introduce the model based on which our analysis will be performed.  In
Section \ref{sec:bbn}, we discuss the BBN constraints on the gravitino
LSP scenario with right-handed sneutrinos.  We will see that the
constraints could be significantly relaxed compared to the case
without right-handed sneutrinos.  Constraints from the structure
formation will be discussed in Section \ref{sec:structure}.
Application of our discussion to the scenario where all the gravitino
dark matter is produced by the decay of other superparticle will be
discussed in Section \ref{sec:superwimp}.  Section
\ref{sec:conclusions} is devoted to conclusions and discussion.

\section{Model Framework}
\label{sec:model}

In this section, we summarize the model.  As we mentioned, we consider
the case where gravitino is the LSP while right-handed sneutrino is
the NLSP.  We assume that neutrino masses are purely Dirac-type, and
the superpotential of the model is written as
\begin{eqnarray}
 W
 =
 W_{\rm MSSM} 
 +
 y_{\nu} \hat{L} \hat{H}_u \hat{\nu}^c_R,
\end{eqnarray}
where $W_{\rm MSSM}$ is the superpotential of the MSSM, $\hat{L} =
(\hat{\nu}_L, \hat{e}_L)$ and $\hat{H}_u=(\hat{H}^+_u, \hat{H}^0_u)$
are left-handed lepton doublet and up-type Higgs doublet,
respectively. In this article, ``hat'' is used for superfields, while
``tilde'' is for superpartners. Generation indices are omitted for
simplicity.  In this model, neutrinos acquire their masses only
through Yukawa interactions as $m_{\nu} = y_{\nu} \langle H^0_{u}
\rangle = y_{\nu} v \sin{\beta}$, where $v \simeq$ 174 GeV is the
vacuum expectation value (VEV) of the standard model Higgs field and
$\tan{\beta} = \langle H^0_u \rangle/\langle H^0_d \rangle$. Thus, the
neutrino Yukawa coupling is determined by the neutrino mass as
\begin{eqnarray}
 y_{\nu} \sin{\beta}
 =
 3.0 \times 10^{-13}
 \times 
 \left(
  \frac{ m^2_{\nu} }{ 2.8 \times 10^{-3} ~{\rm eV}^2 }
 \right)^{1/2}.
\end{eqnarray}
Mass squared differences among neutrinos have already been determined
accurately by neutrino oscillation experiments \cite{K2K,KamLAND}, and
are given by
\begin{eqnarray}
 \left[ \Delta m_{\nu}^2 \right]_{\rm atom}
 \simeq
 2.8 \times 10^{-3}~{\rm eV^2},
 \qquad
 \left[ \Delta m_{\nu}^2 \right]_{\rm solar}
 \simeq
 7.9 \times 10^{-5}~{\rm eV^2}.
 \label{dm}
\end{eqnarray}
In this article, we assume that the spectrum of neutrino masses is
hierarchical, hence the largest neutrino Yukawa coupling is of the
order of $10^{-13}$.  We use $y_{\nu}=3.0 \times 10^{-13}$ for our
numerical analysis.

With right-handed (s)neutrinos, it is also necessary to introduce soft
SUSY breaking terms related to right-handed sneutrinos: right-handed
sneutrino mass terms and tri-linear coupling terms called
$A_\nu$-terms.  Soft SUSY breaking terms relevant to our analysis are
\begin{eqnarray}
 {\cal L}_{\rm SOFT} 
 =
 - M^2_{\tilde{L}}     \tilde{L}^{\dagger} \tilde{L} 
 - M^2_{\tilde{\nu}_R} \tilde{\nu}^{*}_R   \tilde{\nu}_R
 + \left( A_{\nu} \tilde{L} H_u \tilde{\nu}^c_R + {\rm h.c} \right),
\end{eqnarray}
where all breaking parameters, $M_{\tilde{L}}$, $M_{\tilde{\nu}_R}$
and $A_{\nu}$, are defined at the electroweak (EW) scale. We
parametrize $A_{\nu}$ by using the dimensionless constant $a_\nu$ as
\begin{eqnarray}
  A_{\nu} = a_\nu y_{\nu} M_{\tilde{L}}.
  \label{A_nu}
\end{eqnarray}
We adopt gravity-mediated SUSY breaking scenario and, in such a case,
$a_\nu$ is expected to be $O(1)$.  Though the $A_\nu$-term induces the
left-right mixing in the sneutrino mass matrix, the mixing is safely
neglected in the calculation of mass eigenvalues due to the smallness
of neutrino Yukawa coupling constants.  Thus, the masses of sneutrinos
are simply given by
\begin{eqnarray}
 m^2_{\tilde{\nu}_L}
 =
 M^2_{\tilde{L}} + \frac{1}{2} \cos (2 \beta) m^2_Z,
 \quad
 m^2_{\tilde{\nu}_R}
 =
 M^2_{\tilde{\nu}_R},
\label{snumass}
\end{eqnarray}
where $m_Z \simeq$ 91 GeV is the Z boson mass.  In the following
discussion, we assume that all the right-handed sneutrinos are
degenerate in mass for simplicity.

In this article, we consider the gravitino LSP scenario with
right-handed sneutrino NLSP.  In such a case, the next-to-next-the-LSP
(NNLSP) plays an important role in the thermal history of the
universe.  However, there are many possibilities of the NNLSP,
depending on the detail of SUSY breaking scenario.  In our study, we
concentrate on the case that the NNLSP is the lightest neutralino
whose composition is almost Bino. This situation is easily obtained if
we consider the so-called constrained-MSSM type scenario
\cite{Drees:2004jm}. It is not difficult to extend our discussion to
the scenario with other NNLSP candidates.  If Bino is the NNLSP, only
the right-handed sneutrino which is the superpartner of the heaviest
neutrino plays important roles in the Bino decay and other
right-handed sneutrinos are hardly produced in this decay.  Thus,
$m_{\tilde{\nu}_R}$ given in Eq.\ (\ref{snumass}) should be understood
as the mass of the superpartner of the heaviest neutrino.

\section{Constraints from BBN}
\label{sec:bbn}

It is well known that models with the gravitino LSP usually receive
stringent constraints from BBN. In these models, the lightest
superparticle in the MSSM sector (which we call MSSM-LSP) has a long
lifetime, sometimes much longer than one second due to Planck-scale
suppressed interactions. Then, MSSM-LSP produced in the early
universe decays into gravitino by emitting standard-model particles
after BBN starts, which may spoil the success of BBN. In this section,
we show that, when the right-handed sneutrino is the NLSP, the thermal
history of the universe could be significantly altered, resulting in
weaker constraints from BBN.

Now, we consider the thermal history of the universe. In the early
universe when the temperature is higher than the masses of MSSM
particles, all the standard model particles and their superpartners
are in thermal equilibrium. Gravitino and right-handed sneutrinos are,
however, never thermalized due to the weakness of their interactions.
When the temperature becomes as low as $m_{\tilde{B}}/20$
($m_{\tilde{B}}$: mass of Bino-like neutralino), Bino-like neutralino
$\tilde{B}$ decouples from the thermal bath.  Then, Bino-like
neutralino decays into right-handed sneutrino as well as into
gravitino.  After that, right-handed sneutrino decays into gravitino
emitting right-handed neutrino. As can be easily understood, the decay
of right-handed sneutrino is harmless for the BBN scenario, because
only right-handed neutrino is emitted. On the other hand, the decay of
Bino-like neutralino into gravitino affects BBN.

Constraints from BBN including hadronic decay modes are intensively
studied in Refs.\cite{KawKohMor,Jedamzik:2006xz}; according to the
studies, the BBN constraints give the upper bound on $Y_{X} E_{\rm
  vis}$ as a function of $\tau_{X}$, where $X$ stands for a long-lived
but unstable particle, $Y_{X}\equiv [n_{X}/s]_{t\ll \tau_X}$ (with
$n_X$ and $s$ being the number density of $X$ and the entropy density
of the universe, respectively), $E_{\rm vis}$ is the mean energy of
visible particles emitted in the $X$ decay, and $\tau_{X}$ is the
lifetime of $X$. We use the upper bound on $Y_X E_{\rm vis}$ obtained
in Ref.\cite{KawKohMor}. In our analysis, we adopt the line of
$Y_p$(IT) in that work as the constraint from the $p \leftrightarrow
n$ conversion.

\subsection{Decay of Bino-like neutralino}

\begin{figure}[t]
 \begin{center}
  \includegraphics[scale=0.4]{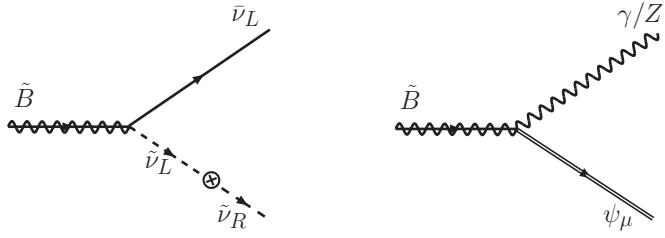}
  \caption{\small Two-body decay diagrams of 
    Bino-like neutralino.}
  \label{fig:bino2bdy}
 \end{center}
\end{figure}

First, we will take a closer look at the decay of the MSSM-LSP, which
is Bino-like neutralino. Main decay modes are the following two-body
decays shown in Fig.\ref{fig:bino2bdy}: $\tilde{B} \rightarrow
\tilde{\nu}_R \bar{\nu}_L$, $\tilde{B} \rightarrow \psi_{\mu} \gamma$,
and $\tilde{B} \rightarrow \psi_{\mu} Z$. Decay widths of these
processes are given by \cite{Feng:2004mt}
\begin{eqnarray}
 \Gamma_{\tilde{B} \rightarrow \tilde{\nu}_R \bar{\nu}_L}
 &=&
 \frac{g_Y^2}{64\pi} m_{\tilde{B}}
 \left[
  \frac{A_\nu v}{m_{\tilde{\nu}_L}^2-m_{\tilde{\nu}_R}^2}
 \right]^2
 \left[
  1-\frac{m^2_{\tilde{\nu}_R}}{m^2_{\tilde{B}}}
 \right],
 \nonumber \\
 \Gamma_{\tilde{B} \rightarrow \psi_{\mu} \gamma}
 &=&
 \frac{\cos^2 \theta_W}{48 \pi M^2_*}
 \frac{m^5_{\tilde{B}}}{m^2_{3/2}}
 \left[
  1 - \frac{m^2_{3/2}}{m^2_{\tilde{B}}}
 \right]^3
 \left[
  1 + 3 \frac{m^2_{3/2}}{m^2_{\tilde{B}}}
 \right],
 \nonumber \\
 \Gamma_{\tilde{B} \rightarrow \psi_{\mu} Z}
 &=&
 \frac{\sin^2 \theta_W}{48 \pi M^2_*}
 \frac{m^5_{\tilde{B}}}{m^2_{3/2}} F
 \left[
  \left(
   1-\frac{m^2_{3/2}}{m^2_{\tilde{B}}}
  \right)^2
  \left(
   1 + 3 \frac{m^2_{3/2}}{m^2_{\tilde{B}}}
  \right)
  - \frac{m^2_Z}{m^2_{\tilde{B}}} G
 \right],
\end{eqnarray}
where $m_{3/2}$ is the gravitino mass and functions $F$ and $G$ are
defined as
\begin{eqnarray}
 &&
 F(m_{\tilde{B}}, m_{3/2}, m_Z)
 =
 \left[
   1 - \left(
     \frac{m_{3/2} + m_Z}{m_{\tilde{B}}}\right)^2
 \right]^{1/2}
  \left[
   1 - \left(\frac{m_{3/2} - m_Z}{m_{\tilde{B}}}\right)^2
  \right]^{1/2},
 \nonumber \\
 &&
 G(m_{\tilde{B}}, m_{3/2}, m_Z)
 =
 3 + \frac{m^3_{3/2}}{m^3_{\tilde{B}}}
 \left( - 12 + \frac{m_{3/2}}{m_{\tilde{B}}} \right)
 + \frac{m^4_Z}{m^4_{\tilde{B}}}
 - \frac{m^2_Z}{m^2_{\tilde{B}}}
 \left( 3 - \frac{m^2_{3/2}}{m_{\tilde{B}}^2} \right).
\end{eqnarray}
Here, $g_Y$ is the U(1)$_Y$ gauge coupling constant, $M_* \simeq 2.4
\times 10^{18}$ is the reduced Planck mass, and $\theta_W$ is the
Weinberg angle. The lifetime of Bino-like neutralino is given by
$\tau^{-1}_{\tilde{B}} = \Gamma_{\tilde{B}} \simeq 2 \Gamma_{\tilde{B}
  \rightarrow \tilde{\nu}_R \bar{\nu}_L} + \Gamma_{\tilde{B}
  \rightarrow \psi_{\mu} \gamma} + \Gamma_{\tilde{B} \rightarrow
  \psi_{\mu} Z}$, where the factor 2 in front of
$\Gamma_{\tilde{B}\rightarrow\tilde{\nu}_R\bar{\nu}_L}$ comes from the
contribution of the CP conjugate final state.

\begin{figure}[t]
 \begin{center}
  \includegraphics[scale=0.5]{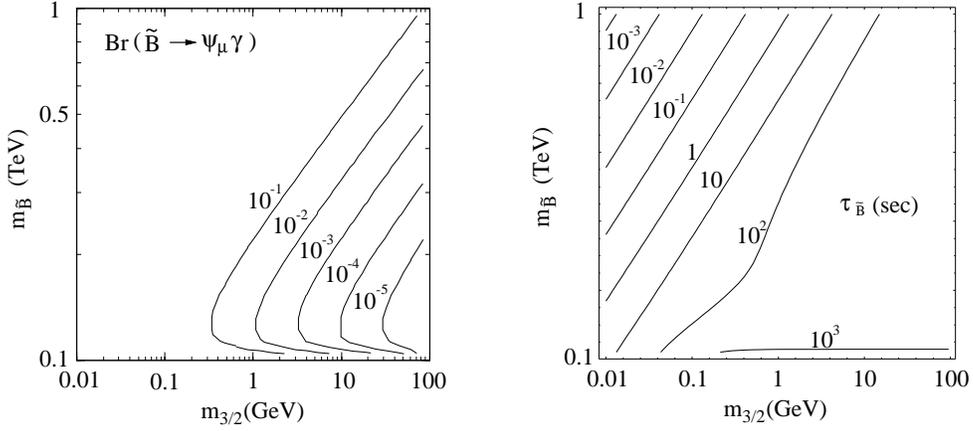}
  \caption{\small Contour plot of the branching ratio Br($\tilde{B}
    \rightarrow \psi_{\mu} \gamma)$ (left figure) and the lifetime of
    $\tilde{B}$ (right figure) on the ($m_{3/2}$, $m_{\small
      \tilde{B}}$) plane: we take $m_{\tilde{\nu}_R} = 100$ GeV,
    $a_{\nu} = 1$, and $m_{\tilde{\nu}_L} = 1.5 m_{\tilde{B}}$ in both
    figures.}
  \label{fig:BgrgammaContourplot}
 \end{center}
\end{figure}

The branching ratio of the mode $\tilde{B} \rightarrow \psi_{\mu}
\gamma$ is shown in Fig.\ref{fig:BgrgammaContourplot} (left figure) on
the ($m_{3/2}$, $m_{\tilde{B}}$) plane, where we take
$m_{\tilde{\nu}_R} = 100$ GeV, $a_{\nu} = 1$, and $m_{\tilde{\nu}_L} =
1.5 m_{\tilde{B}}$.  Importantly, the decay mode
$\tilde{B}\rightarrow\tilde{\nu}_R\bar{\nu}_L$ competes with the mode
$\tilde{B} \rightarrow \psi_{\mu} \gamma$ or it even dominates the
total decay rate when the gravitino mass is larger than 0.1
GeV\footnote{We have checked that $\Gamma_{\tilde{B} \rightarrow
    \psi_{\mu} Z}$ is about one order of magnitude smaller than
  $\Gamma_{\tilde{B} \rightarrow \psi_{\mu} \gamma}$ on the parameter
  region of Fig.\ref{fig:BgrgammaContourplot}.}.  The lifetime of
$\tilde{B}$ is $10^{2-3}$ seconds on most of the parameter region
shown in Fig.\ref{fig:BgrgammaContourplot} (right figure).

Without right-handed sneutrinos, $\tilde{B}\rightarrow\psi_{\mu}\gamma
/Z$ is the main decay mode, and significant amount of visible
particles (including hadrons) are produced.  As a result, the
gravitino mass is strictly constrained as $m_{3/2} \lesssim 0.1$ GeV
for $\tau_{\tilde{B}} \lesssim 1$ second from BBN
\cite{Feng:2004mt}. In our scenario with the right-handed sneutrino
NLSP, however, less visible particles are emitted, though the
Bino-like neutralino is long-lived\footnote{Left-handed neutrinos
  injected by the decay might possibly change the abundance of
  ${}^4$He \cite{Kanzaki}. However, we have checked that the BBN
  constraints on the neutrino injection are much weaker than those on
  hadron injection from three- or four-body decay.}. Therefore,
constraints from BBN is expected to be relaxed.

Next, we consider three- or four-body decay modes of
$\tilde{B}$. Although branching ratios of these processes are much
smaller than 1, they have impacts on the BBN scenario. In particular,
the hadronic decay modes give more severe constraints on the model
than radiative ones via hadro-dissociation and $p \leftrightarrow n$
conversion processes.

\begin{figure}[t]
  \begin{center}
    \includegraphics[scale=0.5]{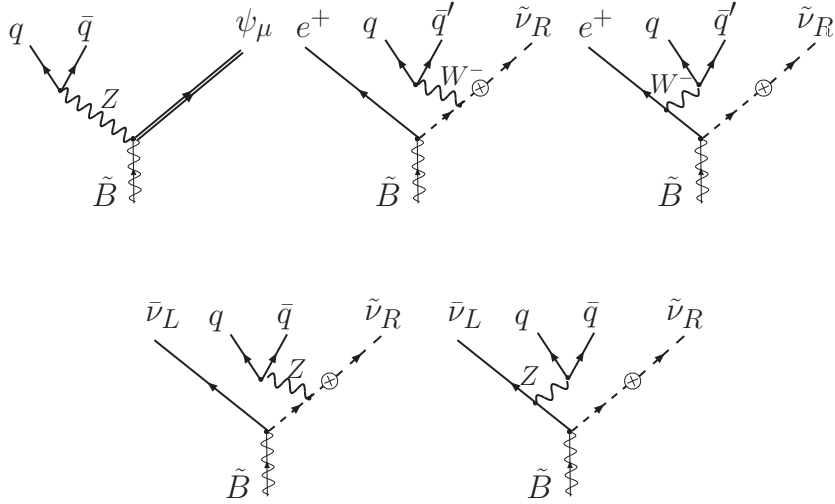}
    \caption{\small Three- or four-body decay diagrams of Bino-like
      neutralino.}
    \label{fig:bino3and4bdy}
  \end{center}
\end{figure}

In our model, three- or four-body decay modes of the Bino-like
neutralino are $\tilde{B} \rightarrow \psi_{\mu} q \bar{q}$,
$\tilde{B} \rightarrow \tilde{\nu}_R e^+_L q \bar{q}^{\prime}$, and
$\tilde{B} \rightarrow \tilde{\nu}_R \bar{\nu}_L q \bar{q}$ shown in
Fig.\ref{fig:bino3and4bdy}, where $q$ and $\bar{q}$ denote quark and
anti-quark, respectively.  Since the hadronic
branching ratio is comparable to radiative one in our scenario, the
hadronic processes give the most severe constaints.  Thus, we
concentrate on hadro-dissociation and $p\leftrightarrow n$ conversion
processes in order to derive BBN constraints.  We calculate $B_{\rm
  had} Y_{\tilde{B}} E_{\rm vis}$ as a function of $\tau_{\tilde{B}}$,
where $B_{\rm had}$ is the hadronic branching ratio. In order to
constrain the model quantitatively, we use the upper bound on $Y_X
E_{\rm vis}$ obtained in Ref.\cite{KawKohMor}\footnote{In some
  parameter regions where the $Z$ boson in $\tilde{B} \rightarrow
  \psi_{\mu} q \bar{q}$ is off-shell, the photo-dissociation effect
  caused by $\tilde{B} \rightarrow \psi_{\mu} \gamma$ might be
  comparable to the hadro-dissociation effect. In such a parameter
  region, however, $\tilde{B} \rightarrow \tilde{\nu}_R \bar{\nu}_L$
  dominates in total decay as seen in
  Fig.\ref{fig:BgrgammaContourplot}. Therefore, the photo-dissociation
  effect is negligible, which is consistent with previous work
  \cite{Feng:2004mt}.}.

The product of the hadronic branching ratio and the visible energy,
which is the mean energy of emitted hadrons from the three- and
four-body decays, is given by
\begin{eqnarray}
 B_{\rm had}E_{\rm vis} 
 &=&
 \frac{1}{\Gamma_{\tilde{B}}}
 \Bigg[
 \sum_q  \left\{
   2 \Gamma_{\tilde{B} \rightarrow \tilde{\nu}_R \bar{\nu}_L q \bar{q}}
   \langle E_{{\rm vis}}^{(\tilde{\nu}_R \bar{\nu}_L q \bar{q})} \rangle
   +
   2 \Gamma_{\tilde{B} \rightarrow \tilde{\nu}_R e^+_L q \bar{q}^{\prime}}
   \langle E_{{\rm vis}}^{(\tilde{\nu}_R e^+_L q \bar{q}^{\prime})} \rangle
 \right\}
 \nonumber \\ & & \qquad
 +
 \Gamma_{\tilde{B} \rightarrow \psi_{\mu} Z} B_{\rm had}^{Z}
  E_{{\rm vis}}^{(Z)}
 \Bigg],
\end{eqnarray}
where the factor 2 in first and second terms are from contributions of
CP conjugate final states, $B^Z_{\rm had} \simeq 0.7$ is the hadronic
branching ratio of $Z$ boson, $\langle E_{\rm vis}^{(\cdots)} \rangle$
is the averaged energy of hadrons emitted in each decay process, and
$E_{\rm vis}^{(Z)}$ is the energy of the $Z$ boson,
\begin{eqnarray}
 E^{(Z)}_{\rm vis}
 =
 \left[
  m^2_Z 
  +
  \frac{m^2_{\tilde{B}}}{4}
  \left(
   1
   -
   2 \frac{m^2_{3/2} + m^2_Z}{m^2_{\tilde{B}}}
   +
   \frac{(m^2_{3/2} - m^2_Z)^2}{m^4_{\tilde{B}}}
  \right)
 \right]^{1/2} .
\end{eqnarray}

\subsection{Constraints}

In order to evaluate $B_{\rm had} Y_{\tilde{B}} E_{\rm vis}$, we have
to determine the primordial abundance of Bino-like neutralino. The
abundance depends highly on parameters in the MSSM sector such as
masses of other superparticles. We use the following formula for
the (would-be) density parameter of $\tilde{B}$ \cite{Feng:2004mt}:
\begin{eqnarray}
 \Omega_{\tilde{B}} h^2 
 =
 C_{\rm model} \times 0.1
 \left[
  \frac{m_{\tilde{B}}}{100~{\rm GeV}} 
 \right]^2,
 \label{eq:omegabinomodel}
\end{eqnarray}
where the additional parameter $C_{\rm model}$ is introduced to take
the model dependence into account: $C_{\rm model} \sim
1$ for the neutralino in the bulk region, $C_{\rm model} \sim 0.1$ for
the co-annihilation or funnel region, and $C_{\rm model} \sim 10$ for
the pure Bino case without co-annihilation.  Then, the yield of the
Bino-like neutralino is given by
\begin{eqnarray}
 Y_{\tilde{B}}
 =
 C_{\rm model} \times
 3.6 \times 10^{-12}
 \frac{m_{\tilde{B}}}{100~{\rm GeV}}. 
\end{eqnarray}

\begin{figure}
 \begin{center}
  \includegraphics[scale=0.45]{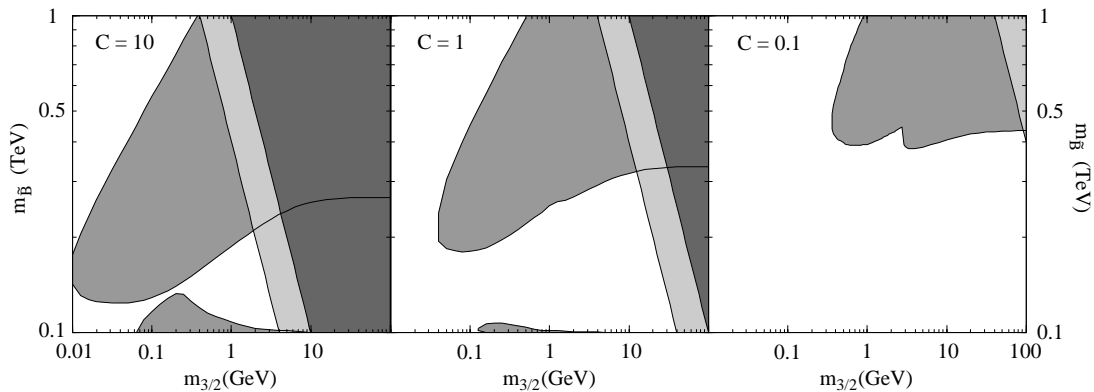}
  \caption{\small Constraints from the BBN on the ($m_{3/2}$,
    $m_{\tilde{B}}$) plane: Parameters are chosen to be
    $m_{\tilde{\nu}_R} = 100$ GeV, $a_{\nu} = 1$, and
    $m_{\tilde{\nu}_L} = 1.5 m_{\tilde{B}}$. We set $C_{\rm model} =
    $10, 1, and 0.1 in the left, middle, and right figures,
    respectively. The middle-shaded regions are ruled out by BBN,
    while dark and light-shaded regions are excluded by the WMAP
    measurement and the structure formation of the universe,
    respectively.}
 \label{fig:MgvsMbinoV2-4}
 \end{center}
\end{figure}

Our numerical results are shown in Fig.\ref{fig:MgvsMbinoV2-4}, where
the BBN constraints are depicted on the ($m_{3/2}$, $m_{\tilde{B}}$)
plane. We take $C_{\rm model} = 10, 1, 0.1$ in the left, middle, and
right figures, respectively. Other parameters are the same as those
used in Fig.\ref{fig:BgrgammaContourplot}. Shaded regions are ruled
out by BBN. As shown in these figures, the constraints are drastically
relaxed compared to those in models without right-handed sneutrinos
\cite{Feng:2004mt}.

As shown in the figures, new allowed region appears; for example, for
$C_{\rm model}=1$, 0.1 GeV $\lesssim m_{3/2} \lesssim$ 40 GeV.  In
that region, $m_{\tilde{B}}$ is bounded from above due to the BBN
constraints from four-body decays, $\tilde{B} \rightarrow
\tilde{\nu}_R e^+_L q \bar{q}^{\prime}$ and $\tilde{B} \rightarrow
\tilde{\nu}_R \bar{\nu}_L q \bar{q}$.  (Notice that the hadronic
branching ratio $B_{\rm had}$ and mean energy $E_{\rm vis}$ are
enhanced when $m_{\tilde{B}}$ is large.)  On the contrary, in the 0.01
GeV $\lesssim m_{3/2} \lesssim$ 0.1 GeV region, Bino-like neutralino
decays mainly into the gravitino through the $\tilde{B} \rightarrow
\psi_{\mu} \gamma$ process with the lifetime $\tau_{\tilde{B}}
\lesssim 1$ second. Since the decay occurs before BBN starts, it does
not affect the BBN scenario. This situation also holds in the usual
gravitino LSP scenario without right-handed sneutrinos, and the same
allowed region can be seen in Ref.\cite{Feng:2004mt}. In the case of
$C_{\rm model} = 10 (0.1)$, the constraints from BBN is more (less)
stringent than the $C_{\rm model} = 1$ case. As a result, the upper
bound on $m_{\tilde{B}}$ becomes smaller (larger).  Results also
depend on left-right mixing angle $\theta_{\tilde{\nu}_L \text{-}
  \tilde{\nu}_R} \equiv |A_{\nu} v
/(m^2_{\tilde{\nu}_L}-m^2_{\tilde{\nu}_R})|$ in the sneutrino mass
matrix.  As one can see in Eq.(7), branching ratio of the mode
$\tilde{B} \rightarrow \psi_{\mu} \gamma / Z$ is more suppressed for
larger $\theta_{\tilde{\nu}_L \text{-} \tilde{\nu}_R}$, which leads to
less stringent constraints from BBN.  On the contrary, when $a_\nu$ in
Eq.(\ref{A_nu}) is much smaller than $1$, BBN constraints are severe
and our constraints become close to those for the case without
right-handed sneutrinos.  We also find that the region $m_{\tilde{B}}
\simeq 100~{\rm GeV}$ and 0.1 GeV $\lesssim m_{3/2} \lesssim$ 1 GeV is
excluded in the left and middle figures.  In these resions, Bino is
almost degenerate with right-handed sneutrino in mass.  As a result,
the process $\tilde{B} \rightarrow \tilde{\nu}_R \bar{\nu}_L$ is
kinematically suppressed and branching ratio of the process $\tilde{B}
\rightarrow \psi_{\mu} Z$ is enhanced.

In addition to the BBN constraints, we also depict other cosmological
bounds in Fig.\ref{fig:MgvsMbinoV2-4}: the gravitino abundance
originating in $\tilde{B}$ must not exceed the value observed in the
WMAP, $\Omega_{\rm DM} h^2 \simeq 0.105$
\cite{Spergel:2006hy}. Gravitino abundance from the decay is given by
$\Omega^{{\rm dec}}_{3/2} = (m_{3/2}/m_{\tilde{B}})~
\Omega_{\tilde{B}}$. Using Eq.(\ref{eq:omegabinomodel}), the
cosmological constraint on the ($m_{3/2}, m_{\tilde{B}}$) plane is
obtained as $m_{\tilde{B}} m_{3/2} < 10^4~{\rm GeV}/C_{\rm model}$,
which is shown as a dark-shaded region in
Fig.\ref{fig:MgvsMbinoV2-4}. This constraint gives the upper bound on
$m_{3/2}$. Another constraint, $\Omega^{\rm dec}_{3/2} <
0.4\Omega_{\rm DM}$, is also depicted as a light-shaded region, which
comes from the structure formation of the universe, which is discussed
in the next section.

\section{Constraints from Structure Formation}
\label{sec:structure}

As shown in the previous section, larger value of $m_{3/2}$ is allowed
compared to the case without right-handed sneutrinos.  In the newly
allowed parameter region, the MSSM-LSP decays mainly into right-handed
sneutrino $\tilde{\nu}_R$, and $\tilde{\nu}_R$ decays into the
gravitino. Since the gravitino is produced with large velocity
dissipation at the late universe, it behaves as a warm dark matter,
and as a result, may affect the structure formation of the
universe. In this section, we consider the constraints from the
structure formation.

\subsection{Decay of right-handed sneutrino}

Once the gravitino is produced from the decay of $\tilde{\nu}_R$, it
is expected to freely stream in the universe and smooth out the (small
scale) primordial density fluctuation.  Since the lifetime of Bino is
much shorter than that of right-handed scneutrino in our scenario, the
free-streaming length of the gravitino is estimated as
\begin{eqnarray}
 \lambda_{\rm FS}
 =
 \int^{t_{\rm EQ}}_{\tau_{\tilde{\nu}_R}} dt
 \frac{v(t)}{a(t)}
 &=&
 \quad
 \frac{2 t_{\rm EQ} u(t_{\rm EQ})}{a(t_{\rm EQ})}
 \left(
  \ln
  \left[
   \frac{1}{u(t_{\rm EQ})} + \sqrt{1 + \frac{1}{u^2(t_{\rm EQ})}}~
  \right]
 \right.
 \nonumber \\
 &&
 \qquad\qquad\qquad~~
 \left.
 -
  \ln
  \left[
   \frac{1}
   {u(\tau_{\tilde{\nu}_R})} + \sqrt{1 + \frac{1}{u^2(\tau_{\tilde{\nu}_R})}}~
  \right]~
 \right),\
\end{eqnarray}
where $v(t)$ is the velocity, $u(t) = p(t)/m_{3/2}$ with $p(t)$ being
the momentum of gravitino, and $a(t)$ is the cosmic scale
factor. Here, the time of the matter-radiation equality and the
lifetime of $\tilde{\nu}_R$ are denoted by $t_{\rm EQ}$ and
$\tau_{\tilde{\nu}_R}$, respectively.

\begin{figure}[t]
  \begin{center}
    \includegraphics[scale=0.35]{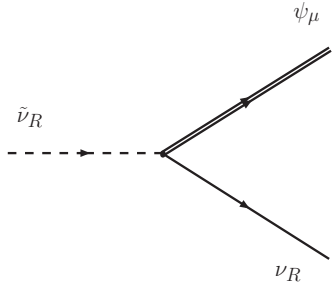}
    \caption{Diagram of the decay of $\tilde{\nu}_R$:
      $\tilde{\nu}_R \rightarrow \psi_{\mu} \nu_R$.}
    \label{fig:snur2bdydecay}
  \end{center}
\end{figure}

Right-handed sneutrino decays into the gravitino through the two-body
decay process $\tilde{\nu}_R \rightarrow \psi_{\mu} \nu_R$ shown in
Fig.\ref{fig:snur2bdydecay}. Notice that three- and four-body decay
processes such as $\tilde{\nu}_R \rightarrow \psi_{\mu} {\nu}_L Z$,
$\tilde{\nu}_R \rightarrow \psi_{\mu} e^-_L W^+$, $\tilde{\nu}_R
\rightarrow \psi_{\mu} {\nu}_L q \bar{q}$, and $\tilde{\nu}_R
\rightarrow \psi_{\mu} e^-_L q\bar{q}'$ are negligible, because these
are strongly suppressed by the small neutrino mass compared to the
two-body decay process. Thus, the decay width of $\tilde{\nu}_R$ is
given by
\begin{eqnarray}
 \Gamma_{\tilde{\nu}_R \rightarrow \psi_{\mu} \nu_R}
=
\tau_{\tilde{\nu}_R}^{-1}
=
 \frac{1}{48 \pi M^2_*} \frac{m^5_{\tilde{\nu}_R}}{m^2_{3/2}}
 \left[
  1 - \frac{m^2_{3/2}}{m^2_{\tilde{\nu}_R}}
 \right]^4.
\end{eqnarray}
The lifetime of right-handed sneutrino turns out to be $10^2 \text{-}
10^8$ seconds with $m_{\tilde{\nu}_R}= 100~{\rm GeV}$ for $0.1\ {\rm
  GeV}\lesssim m_{3/2}\lesssim 100\ {\rm GeV}$.

When $m_{3/2}$ is small enough compared to $m_{\tilde{\nu}_R}$,
$t_{\rm eq} \gg \tau_{\tilde{\nu}_R}$ is satisfied and free-streaming
length is approximately proportional to
$u(\tau_{\tilde{\nu}_R})\tau^{1/2}_{\tilde{\nu}_R}$.  Since
$u(\tau_{\tilde{\nu}_R})^{-1}$ and $\tau^{1/2}_{\tilde{\nu}_R}$ are both
proportional to $m_{3/2}$, $\lambda_{\rm FS}$ becomes independent of
$m_{3/2}$.  With the use of the lifetime obtained in the above
equation, the free-streaming length turns out to be $\lambda_{\rm FS}
\simeq 6$ Mpc when $m_{\tilde{\nu}_R}=100$ GeV unless $m_{3/2}$ is
very close to $m_{\tilde{\nu}_R}$.  This fact indicates that the
component of the dark matter (i.e., gravitino) from sneutrino decay
acts as a warm dark matter (WDM).

In addition to the sneutrino decay, gravitinos are also produced by the
thermal scattering at the reheating epoch after inflation. The
abundance of the gravitino from the scattering process is determined
by the reheating temperature and the gravitino mass
\cite{Moroi:1993mb,Bolz}. Since the gravitino from the scattering is
non-relativistic at the time of the structure formation, it acts as a
cold dark matter (CDM). Thus, we have to consider the constraints from
the structure formation of the universe on the WDM + CDM scenario.

\subsection{Constraints}

Constraints from the structure formation on the WDM + CDM scenario are
studied in recent works \cite{Viel:2005qj, Kaplinghat}. According to
these studies, it turns out that the matter power spectrum has a
step-like decrease around the free-streaming scale of the WDM
component, $k\sim 2\pi/\lambda_{\rm FS}$. This is because only the
power spectrum of the WDM component dumps at $2\pi/\lambda_{\rm FS}$.
The power spectrum is estimated from the observations of the cosmic
microwave background \cite{WMAP1,Spergel:2006hy}, the red shift
surveys of galaxies \cite{Seljak:2004xh}, and so on.  The WDM + CDM
scenario is viable if the step-like decrease is within the uncertainty
of the observed power spectrum.  In this article, we adopt the
following constraint: the power spectrum with the step-like decrease
should be consistent with observational data \cite{WMAP1} at 95\%
confidence level.

In our model, the energy density of dark matter is composed of two
components, $\rho_{\rm DM} = \rho^{\rm dec}_{3/2} + \rho^{\rm
  th}_{3/2}$, where $\rho^{\rm dec}_{3/2}$ and $\rho^{\rm th}_{3/2}$
are the energy densities of gravitino produced by the decay and by the
thermal scattering processes, respectively. Introducing the fraction
of WDM component $f$, we rewrite $\rho_{\rm DM}$ as
\begin{eqnarray}
\rho_{\rm DM} 
= \rho^{\rm dec}_{3/2} + \rho^{\rm th}_{3/2}
= f \rho_{\rm pureWDM} + (1-f) \rho_{\rm pureCDM},
\end{eqnarray}
where $\rho_{\rm pureWDM}$ and $\rho_{\rm pureCDM}$ are the energy
densities of pure WDM and CDM scenario, respectively, where
$\Omega_{\rm pureCDM} h^2=\Omega_{\rm pureWDM} h^2 \simeq 0.1$.  We
consider the adiabatic density fluctuation, then the power spectrum
for the scale $k^{-1}$ is written as
\begin{eqnarray}
  P_{\rm DM}(k) =
  \left[ f P_{\rm pureWDM}^{1/2}(k) + (1-f) P_{\rm pureCDM}^{1/2}(k) \right]^2.
\end{eqnarray}
In order to evaluate the magnitude of the step-like decrease, it is
convenient to define the ratio of the CDM component in the total power
spectrum at $k \gtrsim 2 \pi/\lambda_{\rm FS}$:
\begin{eqnarray}
  r \equiv 
  \frac{P_{\rm DM}(k\gtrsim 2 \pi/\lambda_{\rm FS})}
  {P_{\rm pureCDM}(k \gtrsim 2 \pi/\lambda_{\rm FS})}
  = (1-f)^2. 
\end{eqnarray}
The lower bound on $r$ is obtained from the ratio of the lower and
upper bounds on the observed power spectrum.  In our scenario,
$\lambda_{\rm FS}\simeq 6\ {\rm Mpc}$ and hence $k\simeq 1\ {\rm
  Mpc}^{-1}$.  For such a wavelength, we obtain $r>0.35$ at the 95\%
confidence level \cite{WMAP1}, which leads to $f<0.4$.  In terms of
the density parameter, it indicates $\Omega^{\rm dec}_{3/2}<0.4
\Omega_{\rm DM}$, which gives the upper bound on the gravitino mass as
$m_{3/2}<40$ GeV ($4$ GeV) for $C_{\rm model} = 1$ (10) (light-shaded
regions in Fig.\ref{fig:MgvsMbinoV2-4}).

The constraint does not depend highly on the detail of the
observational data.  In fact, in other recent observations, it is
claimed that the observational error on the power spectrum is about
15\% \cite{Seljak:2004xh, Viel:2005qj}, leading to the constraint as
$f\lesssim 0.2$, which is of the same order of magnitude as the
result above.

\section{Gravitino Dark Matter from Decay}
\label{sec:superwimp}

In our model, it is also possible to realize the scenario in which all
the SuperWIMP (in our case, gravitino) dark matter is produced by the
decay of other superparticle
\cite{FengRajaramanTakayama,Cembranos:2005us}.  In such a scenario,
the SuperWIMP dark matter originates in MSSM-LSP\footnote{Thus, it is
  implicitly assumed that the reheating temperature is low enough in
  order to suppress the gravitino production from the thermal
  scattering at the reheating epoch.}. It is well known that the relic
abundance of the MSSM-LSP explains the observed dark matter abundance
if the MSSM-LSP is stable. Therefore, even if the MSSM-LSP decays into
the SuperWIMP at the late universe, the dark matter abundance is still
explained as far as the mass of the SuperWIMP is of the same order of
that of the MSSM-LSP because of
\begin{eqnarray}
  \Omega_{\rm SuperWIMP} = 
  \frac{m_{\rm SuperWIMP}}{m_{\rm MSSM-LSP}}
  \Omega_{\rm MSSM-LSP}.
  \label{Omega_SWIMP}
\end{eqnarray}

When the SuperWIMP is gravitino, the scenario
receives stringent constraints from BBN
\cite{Feng:2004mt}, because not only gravitino but also visible
particles are emitted in the decay of MSSM-LSP. In order to avoid the
BBN constraints, one may assume that the lifetime of the MSSM-LSP is
much shorter than one second or the MSSM-LSP is highly degenerate with
the gravitino in mass.  However, such possibilities often lead to
fine-tunings of parameters in the model. On the other hand, in our
model, the MSSM-LSP (i.e., Bino-like neutralino) decays mainly into
the right-handed sneutrino NLSP, then right-handed sneutrino decays
into gravitino. The lifetime of $\tilde{B}$ is shorter than that in
model without right-handed sneutrinos, and  the amount of visible
particles emitted in the decay is suppressed. Furthermore, the decay of
$\tilde{\nu}_R$ into gravitino does not produce visible
particles. Thus, the last decay is harmless for BBN, though the
lifetime of sneutrino is much longer than one second. As a result, the
constraints on the scenario are relaxed.

\begin{figure}[t]
  \begin{center}
    \includegraphics[scale=0.65]{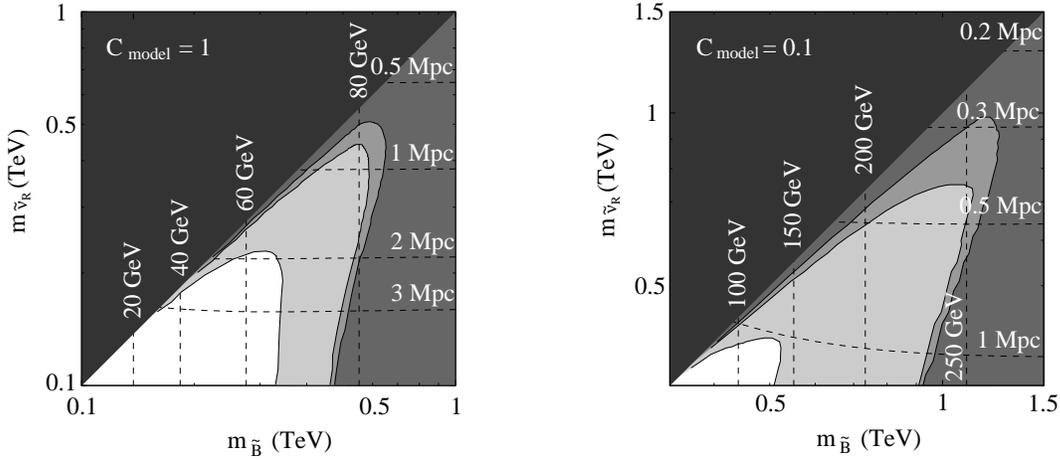}
    \caption{\small BBN constrains to the gravitino dark matter
      scenario on the ($m_{\tilde{B}}$, $m_{\tilde{\nu}_R}$) plane,
      where $m_{\tilde{\nu}_L} = 1.5 m_{\tilde{B}}$.  The region
      $m_{\tilde{\nu}_R}>m_{\tilde{B}}$ is irrelevant because we
      consider the case that $\tilde{\nu}_R$ is the NLSP.  In the left
      figure, we set $C_{\rm model} = 1$, while $C_{\rm model} = 0.1$
      in the right figure. The constraints are shown as light-,
      middle-, and dark-shaded regions, for $a_\nu = $ 1, 3, and 5,
      respectively.}
    \label{fig:SuperWIMP}
  \end{center}
\end{figure}

In Fig.\ref{fig:SuperWIMP}, we show the BBN constraints to the
scenario on the ($m_{\tilde{B}}$, $m_{\tilde{\nu}_R}$) plane, where we
take $m_{\tilde{\nu}_L} = 1.5 m_{\tilde{B}}$. In the left (right)
figure, we set $C_{\rm model} = 1$ (0.1).  Notice that the gravitino
mass is determined by Eqs.(\ref{eq:omegabinomodel}) and
(\ref{Omega_SWIMP}) with $\Omega_{\rm SuperWIMP} \simeq 0.1$. The
excluded regions from BBN are the light-, middle-, and dark-shaded
regions for $a_\nu = $ 1, 3, and 5, respectively. In these figures,
contours of the mass and the free-streaming length of the gravitino
are also depicted.

Since the gravitino is produced in the decay process, it may act as a
warm dark matter.  As a result, the scenario receives the constraint
from the structure formation of the universe.  From observation, the
free-streaming length should be shorter than about 1 Mpc; with such a
bound, we obtain the lower bound on right-handed sneutrino mass as
$m_{\tilde{\nu}_R}\gtrsim 400$ GeV.

As shown in Fig.\ref{fig:SuperWIMP}, when $C_{\rm model} = 1$, only
the small region, $m_{\tilde{B}} \sim 600$ GeV and $m_{\tilde{\nu}_R}
\sim 400$ GeV, is consistent with both of the constraints from BBN and
the structure formation of the universe. The upper bound on
$m_{\tilde{B}}$ comes from the BBN constraints on three- and four-body
decays as discussed in previous sections. On the other hand, the upper
bound on $m_{\tilde{\nu}_R}$ is due to those from the decay of
$\tilde{B}$ into the gravitino.  When $C_{\rm model} = 0.1$, the BBN
constraints are relaxed, because the gravitino mass is larger than
that in the $C_{\rm model} = 1$ case and the decay rate into gravitino
is suppressed. The constraints from the structure formation are also
milder. As a result, a wide range of the parameter space is consistent
with the constraints, which indicates that the scenario with gravitino
dark matter from the MSSM-LSP can be naturally realized in our model
if the annihilation cross section of the MSSM-LSP is large enough.

Recently, it is pointed out that there are various discrepancies
between numerical simulations for the structure formation based on the
CDM scenario and the observations of substructures in galaxies
\cite{satellite, cusp, halo}\footnote{There are also discussions that
  the observations using the gravitational lensing is quite consistent
  with not the WDM scenario but the CDM one \cite{Bradac:2003hy}.}.
Interestingly, a warm dark matter whose free-streaming length is
slightly less than 1 Mpc may solve the discrepancies
\cite{Cembranos:2005us,Bringmann:2007ft,WDM}.  The free-streaming
length as long as $\lambda_{\rm FS} \sim 1.0 \text{-} 0.4 ~{\rm Mpc}$
is suggested as a solution to the missing satellites problem
\cite{satellite}, which can be easily realized in our scenario.  For
cusp problem \cite{cusp}, however, parameter region which solves the
problem receives noticeable change by analysis methods.  For example,
in \cite{Cembranos:2005us}, it is claimed that the cusp and
missing-satellites problem can be solved simultaneously when
$\lambda_{\rm FS}$ is in appropriate range even if the lifetime of the
decaying particle is as short as $\sim 10^5\ {\rm sec}$; then the cusp
problem can be also solved in our scenario.  On the other hand, in
\cite{Bringmann:2007ft}, it is claimed that a simultaneous solution to
both problems are hardly obtained unless the lifetime of the decaying
particle is longer than $\sim 10^{10}\ {\rm sec}$; such a long
lifetime cannot be realized in our scenario.  The detailed analysis of
the small-scale structure problems are out of the scope of this
article and we leave them as future studies.

\section{Conclusions and Discussion}
\label{sec:conclusions}

In this paper, we have studied the cosmological implications of the
gravitino LSP scenario with the right-handed sneutrino NLSP in the
framework where neutrino masses are purely Dirac-type. In the case
that MSSM-LSP is Bino-like neutralino, it mainly decays into the
right-handed sneutrino with the lifetime $\tau_{\tilde{B}} \sim
10^2$-$10^3$ seconds in the wide range of the parameter region. Though
the MSSM-LSP is long-lived, no visible particles are produced in the
leading process, thus constraints from BBN can be relaxed compared to
the case without right-handed sneutrinos. With the quantitative
analysis of the BBN constraints, we have found the new allowed region,
0.1 GeV $\lesssim m_{3/2} \lesssim$ 40 GeV, when $m_{\tilde{\nu}_R} =
$ 100 GeV. In this region, the BBN constraints give the upper bound on
the Bino mass as $m_{\tilde{B}} \lesssim$ 200-400 GeV, which mainly
comes from hadronic four-body decays. On the other hand, the upper
bound on the gravitino mass is given by the constraints from the
structure formation of the universe. In our scenario, some part of the
gravitino is produced by the decay of right-handed sneutrino at the
late universe. As a result, the gravitino freely streams in the
universe and acts as a WDM. The gravitino is also produced from
thermal scattering processes, which acts as a CDM. Taking the CDM
contribution into account, we have considered the constraints on the
WDM + CDM scenario from the observations of (small scale) structure
formation, and finally found the upper bound on the gravitino mass.

So far, we have concentrated on the case with Bino MSSM-LSP.  Another
well-motivated candidate for the MSSM-LSP is the lighter stau
$\tilde{\tau}_1$.  With $\tilde{\tau}_1$-NNLSP, the main decay mode of
$\tilde{\tau}_1$ is $\tilde{\tau}_1 \rightarrow \tilde{\nu}_R
W$\footnote {If this process is kinematically forbidden,
  $\tilde{\tau}_1$ mainly decays as $\tilde{\tau}_1 \rightarrow
  \psi_{\mu} \tau$ and the right-handed sneutrino plays no significant
  role.  Constrraints on such a scenario are already analyzed in
  \cite{Feng:2004mt,Kawasaki:2007xb}, where the upper bound on the
  gravitino mass is given as $m_{3/2} \lesssim 10~{\rm GeV}$.}.  We
have also analyzed this case and derived upper bound on
$m_{\tilde{\tau}_1}$; we found $m_{\tilde{\tau}_1}\lesssim 400~{\rm
  GeV}$ for $m_{\tilde{\nu}_R} = 100~{\rm GeV}$ and $1~{\rm GeV}
\lesssim m_{3/2} < 100~{\rm GeV}$.  Here we take $U^2_{1L}=0.1$, where
$U_{1L}$ is left-handed stau component in lighter stau.  In addition,
the lifetime of lighter stau is 1-$10^2$ seconds in the allowed
region.  Thus, the lighter stau may be seen as a long-lived charged
track in future colliders.

We have also discussed the possibility to realize the scenario where
gravitino dark matter is produced from the decay of other
superparticle. In the scenario, it is postulated that the gravitino
dark matter originating in the Bino-like neutralino accounts for the
total dark matter abundance. Considering the mass parameter region
$m_{\tilde{\nu}_R} = $ 400 GeV-1 TeV, we have found that the
free-streaming length is $\lambda_{\rm FS} = $ 1-0.3 Mpc, which allows
to solve the small scale structure problems of galaxies.  As in the
case above, BBN constraints give the upper bound on the Bino mass as
$m_{\tilde{B}} \lesssim$ 600 GeV-1 TeV, which corresponds to $m_{3/2}
\lesssim$ 80-250 GeV. As a result, all superparticle masses are within
100 GeV-1 TeV, thus the gravitino dark matter scenario in our
framework seems to be natural.

\section*{Acknowledgments}

This work was supported in part by Research Fellowships of the Japan
Society for the Promotion of Science for Young Scientists (K.I.), and
by the Grant-in-Aid for Scientific Research from the Ministry of
Education, Science, Sports, and Culture of Japan, No.\ 19540255
(T.M.).


\begin{thebibliography}{99}

\bibitem{WMAP1}
    D.~N.~Spergel {\it et al.},
    Astrophys.\ J.\ Suppl.\  {\bf 148}, 175 (2003).

\bibitem{Seljak:2004xh}
    U.~Seljak {\it et al.},
    Phys.\ Rev.\ D {\bf 71}, 103515 (2005).

\bibitem{Spergel:2006hy}
  D.~N.~Spergel {\it et al.}  [WMAP Collaboration],
  Astrophys.\ J.\ Suppl.\  {\bf 170}, 377 (2007).

\bibitem{ReviewDM}
  See, for example, 
  G.~Jungman, M.~Kamionkowski and K.~Griest,
  Phys.\ Rept.\  {\bf 267}, 195 (1996);
  G.~Bertone, D.~Hooper and J.~Silk,
  Phys.\ Rept.\  {\bf 405}, 279 (2005).

\bibitem{AsakaIshiwataMoroi}
  T.~Asaka, K.~Ishiwata and T.~Moroi,
  Phys.\ Rev.\  D {\bf 73}, 051301 (2006);
  Phys.\ Rev.\  D {\bf 75}, 065001 (2007).

\bibitem{Moroi:1993mb}
  T.~Moroi, H.~Murayama and M.~Yamaguchi,
  Phys.\ Lett.\  B {\bf 303}, 289 (1993).

\bibitem{FengRajaramanTakayama}
  J.~L.~Feng, A.~Rajaraman and F.~Takayama,
  Phys.\ Rev.\ Lett.\  {\bf 91}, 011302 (2003);
  Phys.\ Rev.\  D {\bf 68}, 063504 (2003);
  L.~Roszkowski, R.~Ruiz de Austri and K.~Y.~Choi,
  JHEP {\bf 0508}, 080 (2005);
  D.~G.~Cerdeno, K.~Y.~Choi, K.~Jedamzik, L.~Roszkowski and R.~Ruiz de Austri,
  JCAP {\bf 0606}, 005 (2006).


\bibitem{Ellis:2003dn}
  J.~R.~Ellis, K.~A.~Olive, Y.~Santoso and V.~C.~Spanos,
  Phys.\ Lett.\  B {\bf 588}, 7 (2004).

\bibitem{K2K}
    E.~Aliu {\it et al.},
    Phys.\ Rev.\ Lett.\  {\bf 94}, 081802 (2005).

\bibitem{KamLAND}
    T.~Araki {\it et al.},
    Phys.\ Rev.\ Lett.\  {\bf 94}, 081801 (2005).

\bibitem{Drees:2004jm}
  See, for example, M.~Drees, R.~Godbole and P.~Roy,
  ``Theory and phenomenology of sparticles: 
  An account of four-dimensional N=1
  supersymmetry in high energy physics,''
  {\it  Hackensack, USA: World Scientific (2004) 555 p}

\bibitem{KawKohMor}
    M.~Kawasaki, K.~Kohri and T.~Moroi,
    Phys.\ Lett.\ B {\bf 625}, 7 (2005);
    Phys.\ Rev.\ D {\bf 71}, 083502 (2005).

\bibitem{Jedamzik:2006xz}
  K.~Jedamzik,
  Phys.\ Rev.\  D {\bf 74}, 103509 (2006)

\bibitem{Feng:2004mt}
  J.~L.~Feng, S.~f.~Su and F.~Takayama,
  Phys.\ Rev.\  D {\bf 70}, 063514 (2004);
  Phys.\ Rev.\  D {\bf 70}, 075019 (2004).

\bibitem{Kanzaki}
  T.~Kanzaki, M.~Kawasaki, K.~Kohri and T.~Moroi,
  Phys.\ Rev.\  D {\bf 75}, 025011 (2007);
  T.~Kanzaki, M.~Kawasaki, K.~Kohri and T.~Moroi,
  arXiv:0705.1200 [hep-ph].

\bibitem{Bolz}
  M.~Bolz, A.~Brandenburg and W.~Buchmuller,
  Nucl.\ Phys.\  B {\bf 606}, 518 (2001).

\bibitem{Viel:2005qj}
  M.~Viel, J.~Lesgourgues, M.~G.~Haehnelt, S.~Matarrese and A.~Riotto,
  Phys.\ Rev.\  D {\bf 71}, 063534 (2005).

\bibitem{Kaplinghat}
  M.~Kaplinghat,
  Phys.\ Rev.\  D {\bf 72}, 063510 (2005).

\bibitem{Cembranos:2005us}
  J.~A.~R.~Cembranos, J.~L.~Feng, A.~Rajaraman and F.~Takayama,
  Phys.\ Rev.\ Lett.\  {\bf 95}, 181301 (2005).

\bibitem{satellite}
  A.~A.~Klypin, A.~V.~Kravtsov, O.~Valenzuela and F.~Prada,
  Astrophys.\ J.\  {\bf 522}, 82 (1999);
  A.~R.~Zentner and J.~S.~Bullock,
  Astrophys.\ J.\  {\bf 598}, 49 (2003).

\bibitem{cusp}
  B.~Moore,
  Nature {\bf 370}, 629 (1994);
  R.~A.~Flores and J.~R.~Primack,
  Astrophys.\ J.\  {\bf 427}, L1 (1994);
  J.~J.~Binney and N.~W.~Evans,
  Mon.\ Not.\ Roy.\ Astron.\ Soc.\  {\bf 327}, L27 (2001);
  A.~R.~Zentner and J.~S.~Bullock,
  Phys.\ Rev.\  D {\bf 66}, 043003 (2002);
  J.~D.~Simon, A.~D.~Bolatto, A.~Leroy, L.~Blitz and E.~L.~Gates,
  Astrophys.\ J.\  {\bf 621}, 757 (2005).

\bibitem{halo}
  J.~F.~Navarro, C.~S.~Frenk and S.~D.~M.~White,
  Astrophys.\ J.\  {\bf 462}, 563 (1996);
  Astrophys.\ J.\  {\bf 490}, 493 (1997);
  B.~Moore, S.~Ghigna, F.~Governato, G.~Lake, T.~Quinn, J.~Stadel 
  and P.~Tozzi,
  Astrophys.\ J.\  {\bf 524} (1999) L19.

\bibitem{Bradac:2003hy}
  M.~Bradac, P.~Schneider, M.~Lombardi, M.~Steinmetz, 
  L.~V.~E.~Koopmans and J.~F.~Navarro,
  Astron.\ Astrophys.\  {\bf 423}, 797 (2004).

\bibitem{Bringmann:2007ft}
  T.~Bringmann, F.~Borzumati and P.~Ullio,
  arXiv:hep-ph/0701007.

\bibitem{WDM}
  D.~Hooper, M.~Kaplinghat, L.~E.~Strigari and K.~M.~Zurek,
  Phys.\ Rev.\  D {\bf 76}, 103515 (2007);
  R.~Kitano and I.~Low,
  arXiv:hep-ph/0503112;
  R.~B.~Metcalf,
  arXiv:astro-ph/0407298;
  J.~L.~Feng,
  J.\ Phys.\ G {\bf 32}, R1 (2006);
  F.~Wang and J.~M.~Yang,
  Eur.\ Phys.\ J.\  C {\bf 38}, 129 (2004);
  K.~Abazajian and S.~M.~Koushiappas,
  Phys.\ Rev.\  D {\bf 74}, 023527 (2006).
  L.~Hui,
  Phys.\ Rev.\ Lett.\  {\bf 86}, 3467 (2001);
  J.~Hisano, K.~Kohri and M.~M.~Nojiri,
  Phys.\ Lett.\  B {\bf 505}, 169 (2001).

\bibitem{Kawasaki:2007xb}
  M.~Kawasaki, K.~Kohri and T.~Moroi,
  Phys.\ Lett.\  B {\bf 649}, 436 (2007).

\end{thebibliography}
\end{document}